\documentstyle[epsfig,12pt]{article}
\title
{"FREE" CONSTITUENT QUARKS AND DILEPTON PRODUCTION IN HEAVY ION COLLISIONS 
$^*$.}

\author{O.D. CHERNAVSKAYA, E.L. FEINBERG, I.I. ROYZEN}
\begin{document}
\maketitle
\centerline{\em{Lebedev Physical Institute of RAN, Leninski prospect 53,
117333 Moscow,Russia}}
\centerline{e-mail: $<chernav@lpi.ru>,\quad<feinberg@lpi.ru>,\quad
<royzen@lpi.ru>$}
%\addtolength{\baselineskip}{6pt}
\begin{abstract}
An approach is suggested, invoking vitally the notion of constituent massive
quarks (valons) which can survive and propagate rather than hadrons (except of
pions) within the hot and dense matter formed below the chiral transition
temperature in course of the heavy ion collisions at high energies.
This approach is shown to be quite good for
description of the experimentally observed excess in dilepton
yield at masses 250 MeV $\leq\,M_{ee}\,\leq$ 700 MeV over the prompt resonance
decay mechanism (CERES cocktail) predictions. In certain aspects, it looks to be
even more successful, than the conventional approaches: it seems to match
the data somewhat better at dilepton masses before the two-pion threshold and
before the $\rho$-meson peak as well as at higher dilepton masses (beyond the
$\phi$-meson one). The approach implies no specific assumptions on the equation
of state (EOS)
or peculiarities of phase transitions in the expanding nuclear matter.
\end{abstract} \vspace{4,5cm} ---------------------------------\\
$^*$This work is supported by the Russian Foundation for Basic Researches,
grants No.'s 96-15-96798 and 00-02-17250.

\newpage

INTRODUCTION                \\

Few years ago, the experimental evidence has been obtained \cite{exp}
that $e^+ e^-$ pairs (below referred as dileptons) with the invariant masses
250 MeV $\leq\,M\leq$ 700 MeV produced in course of heavy ion collisions at
high energies are by far more numerous (up to a factor about $5\div7$), than
what could be predicted by direct summing up the contributions of known mesonic
resonance decays (CERES cocktail), although the similar treatment of dilepton
yield in the proton-nucleus collisions was quite successful. Since then, many
attempts have been made\cite{Rapp} to put forward a reasonable theoretical
explanation of such a distinction.  Generally, the most of these attempts were
based on the thermodynamical approach \cite{EF1} supplemented by some
assumptions on the kinetics and in-medium properties of hadronic resonances
(changes in their masses and widths \cite{Rapp,Fri}) within the hot and dense
matter (fireball) which was formed in course of the heavy ion collisions at
high energies.  It was demonstrated \cite{Rapp} that under a proper choice of
resonance modifications (predominantly, of the $\rho$-meson width) a seemingly
satisfactory agreement between the above experimental data and their
theoretical treatment can be achieved.  Unfortunately, the relevant models
inevitably suffer from the well known underlying ambiguities - first of all,
from the large extent of freedom in choice of EOS and of
the in-medium particle mass operator.  That is why their predictions are not
undeniable and why the elaboration of some alternative approaches seems to be
not out of place.

The approach we are to discuss below is based on a microscopic picture
of the hot fireball evolution which necessarily implies the essential role of
the massive constituent quarks \cite{EF2}(following R.  Hwa \cite{Hwa},
we call them below as "valons") at the certain
stage of evolution.  The notion of valon was quite fruitfully exploited at the
early age of quark model and was almost forgotten after constructing the
elegant QCD which makes it possible to deduce and predict many phenomena in
terms of current (point-like) quarks and gluons. The attempts to embed valons
rigorously into the framework of QCD as some quazibound color states of quarks
and gluons were not proved to be successful \cite{Wil}, but, being physically
very attractive, the notion of valon was exploited nevertheless for giving the
qualitative motivations in favor of one or another
statement. Among them, the attempts should be mentioned, first of all, to
distinguish between the hadronic breakdown temperature \cite{Hag}
and the chiral symmetry restoration temperature by consideration either two
successive phase transitions \cite{Kal,Shur}
or gradual valonic mass decrease as the temperature rises, being above
the former one \cite{Shap}.

The more detailed attempts to incorporate the valons as certain
phenomenological entities has been made \cite{dptm, Cher} within the framework
of the bag-model EOS of the nuclear matter. Two first order phase transitions
were considered (instead of one within the conventional models) in course of
the fireball expansion:  first - from the short QGP phase to the intermediate
one (chiral symmetry breaking at $T_{ch}\simeq$ 200 MeV) which is rather short
too - it lasts until quick cooling down to the Hagedorn temperature,
$T_H\simeq$ 140 MeV, is completed, and second - long and nearly
isothermic transition (which is referred as mixed pion-valonic state by
analogy with the conventional mixed QGP-hadronic state)
from this phase to the short hadronic phase which ends by freeze-out at
slightly lower temperature, $T_f\simeq$ 120 MeV.

Being undoubtedly different, all these models (including the conventional
ones) show up one common feature: the (mixed) phase preceding the color
confinement lasts much longer, than the other ones, irrespectively of
peculiarities of a specific model.
That is why one can reasonably believe that the substantial deceleration of
expansion and corresponding prolongation of the pion-valonic phase
is inherent not only in all versions of bag model - this
qualitative effect is, most probably, the general and inevitable consequence of
the necessity to meet the color confinement at low densities.  In what follows
we keep this pattern in mind as a guideline.  Thus, within our approach, just
the pion-valonic phase of expansion (not the QGP one) is expected to be
responsible for the "extra dileptons" with low masses (over the CERES cocktail
sample) seen at SPS:  they can be really produced during this long phase in
course of numerous successive collisions of particles within fireball.  Being
quite short, the hadronic phase can provide the resonance background (CERES
cocktail) only.

The physical meaning of these phases in fireball evolution seems quite
transparent \cite{EF2} irrespectively of whether sharp or soft phase
transitions take place as well as of the specific time profile of fireball
temperature or of some other model-dependent peculiarities: the chiral symmetry
breaking (restoration) and color confinement (deconfinement) are assumed to
happen under essentially different thermodynamic conditions.
The valon can be thought of as a quaziparticle in a sense that
it absorbs the most part of strong color interaction to form (within a suitable
range of temperatures and densities) the nearly ideal (color-screened) valonic
gas that is equivalent in its physical manifestations to the gas of strongly
interacting conventional (free) hadrons or QCD (point-like) quarks and gluons.
One can deal with either of the above representations, but of these two
options, the former is obviously by far more comfortable for the theoretical
treatment.  Indeed, within a medium of a density about the nucleon one (in
which the nucleons bodies themselves would occupy the entire volume), the
"equivalent" set of valon bodies (whose radius is supposed \cite{Anis,AQM} to
be about three times smaller, than the nucleon one) would occupy about 10 \% of
the volume only.  Therefore, even at the noticeably higher densities (say, at
the density that is assumed to appear at the chiral phase transition - about
twice as high as the nucleon one or about four times higher, than the nucleus
density), a gaseous approach to the treatment of the valonic matter seems still
reasonable. As usual, one has to pay for this simplification: a poorly
determined entity - the cross section of valon-valon
interaction - enters the relevant formulae inevitably.  Below, a
semi-quantitative approximation is suggested which makes it possible to
overcome this unpleasant obstacle.

It is worthy to note that because of what is said above, the precise and
complicated calculations are unnecessary even at those points where they
really could be performed, when the problem we are interested in is considered.
That is why the rather crude approximations we shall exploit below are
suitable.\\

GENERAL DESCRIPTION OF THE APPROACH\\

The following picture of fireball evolution is adopted:

  *  After cooling down till the temperature $T\,=\,T_{ch}$, QGP
     fireball undergoes the rather quick phase transition\footnote{Nothing,
     except of temperature and baryon density prevents the point-like
     quarks and antiquarks to become "dressed" (i.e., to become the valons and
     antivalons, $Q$ and $\bar Q$).  That is why the valons are reasonably
     expected to appear at the proper conditions almost instantly.}
     which results in formation of two-component (valonic and pionic)
     quaziideal gases in the equilibrium state\footnote{The pions are the
     only hadrons that have a good chance to survive within the medium at this
     stage because the binding energy of valons coupled to form a pion,
     $2\,m_Q\,-\,m_{\pi}\simeq\,500$ MeV, substantially
     exceeds the temperature.  Although the time of establishing the chemical
     equilibrium within fireball is, most probably, somewhat longer, than
     that of the chiral phase transition itself, it is comparatively short too
     \cite{Heinz}.}; the relative content of each species is regulated by the
     detailed balancing principle; the only interaction is accounted that
     converts valons $Q\bar Q$ into pions and vice versa.

 **  This state is maintained sufficiently long
     for producing dileptons numerously via the successive particle
     interactions. As for "macroscopic" patterns (longitudinal and transverse
     flows), they are subjected, as usual, to the relativistic thermo- and
     hydrodynamical treatment. What we need here of all that, is the estimate of
     duration of the pion-valonic phase. So long as, in the end, the
     necessity of color confinement at low densities is motivated above to be
     responsible for its prolongation, no reasons seem to be put forward for
     appearance of an essential difference in this respect between the
     suggested approach and conventional or bag model treatments.  That is why
     the correspondent estimates given by the latter ones \cite{dptm,Cher,Sto}
     will be quoted for orientation.

***  The mesonic resonances are expected to be nearly melted \cite{Ele} over
     almost the entire duration of the pion-valonic phase in a sense that their
     effective widths are crucially influenced by the inverse mean free time
     $\bar t ^{-1}$ which is undoubtedly larger, than the relevant intrinsic
     widths $\Gamma_i$ \footnote{Since $T\,>\,T_H$, the mean free time
     (path) must be shorter or about 1 fm (the radius of valonic confinement
     or of hadronization); the effective width
     of each resonance thus is decreasing along with evolution of the
     intermediate phase to become finally $(\bar t^{-1}\,+\,\Gamma_i)\,\simeq$
     200 MeV + $\Gamma_i$.}. That is why dileptons produced under such
     conditions in the reactions $\pi^+ \pi^- \to e^+ e^- $ and $Q\bar Q\,\to
     e^+ e^-$ are to be treated reasonably as a kind of non-resonance
     multi-collision (transport) contribution (just what we are here to deal
     with) which should be added to dileptons originated from the
     mesonic resonance "normal" decays (CERES cocktail) at the final stage of
     expansion (when these resonances can survive. The heavier are the colliding
     nuclei and the higher is centrality of an observed collision, the more
     numerous should be these transport dileptons\footnote{The dileptons
     coming from the processes of bremsstrahlung type do not practically
     contribute to the part of dilepton mass spectrum under discussion. Indeed,
     typical masses of bremsstrahlung virtual photons emitted by pions or $u$-
     and $d$-valons are wittingly lower, than $m_{\pi}$ or $m_Q$, respectively,
     thus being, in any case, lower, than 300 MeV (in fact, the photons of
     much lower masses are noticeable only). As for heavier valons, yes, they
     could emit such a heavy photons (especially, $c,b$ and $t$ ones), but their
     relative concentration itself is by far too low and one can undoubtedly
     disregard the relevant dilepton yield.}.

  The general strategy of calculations looks as follows:

1. The total numbers of pions, $N_{\pi}$, and valons, $N_Q$ and $N_{\bar Q}$,
within fireball are linked by using the detailed balancing principle.

2. The rate of reactions $\pi^+ \pi^-\to\,e^+ e^-$ and $Q\bar Q\,\to\,e^+ e^-$
is estimated as a function of $M$ and
$N_{\pi}$. Being multiplied by the entire duration of
fireball expansion from the temperature $T_{ch}$ to the temperature $T_f$, it
gives the total yield of dileptons produced via pionic and valonic collisions.

3. $N_{\pi}$ is linked to $N_{ch}$, the total number of charged hadrons coming
from the fireball after freeze-out, and the general formula is adopted for
comparing the results of calculations and experimental data.

4. The obtained results are confronted to the available data for production of
the low mass dileptons, as well as to the results of some conventional
approaches.\\

 DESIGNATIONS \\

\noindent $\frac{d\nu_Q ^w}{dt}$ is the rate of the "white"  $Q\bar
Q$ collisions in which colors of $Q$ and $\bar Q$ are cancelled to produce a
color singlet state.\\ $\frac{d\nu_{\pi}}{dt}$ is the rate of $\pi^+ \pi^-$
and $\pi^0 \pi$ collisions.\\ $\frac{d\nu_Q ^{wo}}{dt}$ and
$\frac{d\nu_{\pi}^o}{dt}$ are the rates of collisions selected from the above
ones in which the particles involved have necessarily the opposite (and non-
zero)
electric charges.  \\ $\lambda\,=\,[\frac{N_Q}{N_{\bar
Q}}]^{1/2}\,\simeq\,exp{\,\frac{\mu_Q}{T}}$, $\mu_Q$ being the chemical
potential of $u$- and $d$-valons, denotes the valon fugacity.\\
$b\,=\,\frac{N_{\pi}}{N_Q +\,N_{\bar Q}}$.\\ $\bar t$ and $\tau$ denote the
mean free time of  particles within fireball and the duration of fireball
cooling from the temperature $T_{ch}$ to the temperature $T_H$, respectively.
\\

 CALCULATIONS\\

1. Being averaged over the particle distributions, the detailed
balancing principle reads:  \begin{equation} \nu_Q ^w (T)\,\overline
{\Omega_{\pi}}(T)\,\simeq\,\nu_{\pi}(T)\,\overline{\Omega_Q}(T)
\end{equation} where $\overline{\Omega_i}$ are the mean values of the
corresponding final state phase spaces. Below, the binary reactions are to be
considered only because 2 $\to$ 4 reactions are substantially suppressed by
scarcity of the typical thermal final state phase space at $T\,<\,T_{ch}$, and 3
(or more) $\to\, anything$ reactions are rather rare events at the typical
particle densities under consideration (however, see the discussion below).
Besides, we restrict ourselves, for a while, with the two lightest
flavors ($N_f=\,2$) because of the low concentration of $s$-quarks: their
number is believed to be about \cite{Sol}
$(0,25\,\div\,0,5)\,exp[(\,m_{u,d}-\,m_s)/T]\,\simeq$ 10 \% of the number of
$(\bar u\,+\,\bar d)$-quarks and thus the relevant corrections are obviously
within the very accuracy of the suggested approach.

Since each antiquark of a certain color and flavor can encounter with the same
probability $\lambda^2$ quarks and 1 antiquark ($2N_c$ species of each of
them) and $b$ pions for each of them, of which only 2 species are
suitable to build a color singlet state, \begin{equation} d\nu_Q^w
(T)\,=\,\frac{\lambda^2\,N_{\bar
Q}}{(\lambda^2\,+\,1)(1\,+\,b)N_c}\,\frac{dt}{\bar t (T)}\end{equation} Quite
similarly, a $\pi^0$-meson encounters another $\pi$-meson with the
probability $\frac{1}{(1\,+\,b^{-1})}\,\frac{dt}{\bar t}$, the total rate of
$\pi^0 \pi$ collisions being, therefore,
$\frac{2N_{\pi}}{9(1\,+\,b^{-1})}\,\frac{dt}{\bar t}$  ($\pi^0
\pi^{\pm}$ collisions) plus $\frac{N_{pi}}{18(1\,+\,b^{-1})}\,\frac{dt}{\bar t}$
 $\pi^0 \pi^0$ collisions); the rate of $\pi^+ \pi^-$ collisions is
obviously $\frac{N_{\pi}}{9(1\,+\,b^{-1})}\,\frac{dt}{\bar t}$. Of course,
$\pi^+\pi^+$ and $\pi^-\pi^-$ collisions are out of the game in the detailed
balancing principle equation (within the above approximation), since they never
result in a two-valonic final state.  Thus, for the total rate of $\pi
\pi$ collisions to be accounted one gets
\begin{equation}
d\nu_{\pi}(T)\,=\,\frac{7}{18}\,\frac{b\,N_{\pi}}{1\,+\,b}\end{equation} The
valonic and pionic phase spaces are \begin{equation} \overline{\Omega_Q}
(T)\,\simeq\,4(2S_Q+\,1)^2 N_c\,\overline{p_Q^2} (T) \quad \mbox{and} \quad
\overline {\Omega_{\pi}}(T)\,\simeq\,(2I_{\pi}+\,1)^2\overline{p^2_{\pi}}
(T),\end{equation} respectively, where $S_Q$ is the valonic spin and $I_{\pi}$
is the pionic isospin, $p_i$ are the valonic and pionic momenta, and $N_c$
stands here instead of $N_c^2$, since only the color singlet part of the total
phase space of two valons is allowed for. The straightforward averaging over
the Boltzmann distribution gives for a particle of the mass $m$ the mean value
of its energy squared $m$ $$\overline{E^2
(m,T)}\,=
\,T^2\,[\,3\frac{\frac{m}{T}\,K_1(\frac{m}{T})}{K_2(\frac{m}{T})}\,+\,12\,+\,
\frac{m^2}{T^2}\,]$$
where $K_{1,2}$ are the corresponding Bessel functions. The CMS value of
$\overline{p^2_{\pi}}$ ($\overline{p^2_Q}$) of each particle in the pionic
(valonic) final state is obtained obviously by insertion into this expression
$m\,=\,m_Q$ ($m\,=\,m_{\pi})$ and subtraction $m^2_{\pi}$ ($m^2_Q$).  Within
the temperature range we are interested in, the ratio of these values varies
slowly and it (namely, the mean value of pionic momentum squared to that of the
valonic one) equals to $\simeq$ 2 at the temperature $\bar T\,\simeq\,160$ MeV
which will be exploited in what follows as certain effective mean temperature
instead of the current one. Making use of the above designations and combining
eq's. (1) - (4), we obtain \begin{equation}
b\,\simeq\,0,6\,\frac{\lambda}{(\lambda^2\,+\,1)}\,\leq\,0,3 \end{equation}
Since the fraction of "big" pions (as compared to "small" valons) is relatively
small (at the reasonable value of $\lambda$, $\lambda\,\simeq\,\sqrt{3}$, that
refers to $\mu_Q \simeq $ 80 MeV, one has $b\,\simeq $
0,24), the motivation in favor of applicability of the gaseous approximation
given above for the purely valonic medium remains valid.  It is worthy
to note also that this chemically equilibrium ratio of valons and pions
corresponds to what would be obtained, if they were considered as being the
ideal non-interacting gases.  This fact enables, at least, to be sure that the
rather crude and idealized gaseous approach to the problem under consideration
is not controversial.

2. The rates of "white" collisions with zero total electric charge
which can produce dileptons via the virtual photon intermediate
state are estimated quite similarly:  \begin{equation}
d\nu_Q^o\,=\,0,5\,d\nu_Q \quad \mbox{and} \quad
d\nu^o_{\pi}\,=\,\frac{b}{9(1\,+\,b)}\,\frac{dt}{\bar t}\end{equation} It is
easy to check that the probability $dW/dM$ of a two particle collision with the
invariant mass (the total energy in their CMS) $M$ is (in  the ideal Boltzmann
gas approximation\footnote{There is no reason to refine it by taking into 
account
the Bose and Fermi statistics of pions and valons, respectively, in face of the
low accuracy and specific kinematical selection (see below) of the available
experimental data.}):  \begin{equation}
\frac{dW}{dM}\,=\,\frac{M}{8}\,\frac{\int_{M}^{\infty} e^{-\xi /T}d\xi
\int_{0}^{\sqrt{(\xi ^2-\,M^2)(1\,-\frac{4m_i^2}{M^2})}}(\xi ^2-\,\eta
^2)d\eta}{\lbrack\int_{0}^{\infty} p^2 e^{-\sqrt{p^2+\,m_i^2}/T}dp\rbrack
^2}\end{equation} where $i\,=\,Q,\pi$, $\,\xi$ and $\eta$ denote the sum and
difference of colliding particle energies, respectively\footnote{The nominator
here is nothing else than the product of the momentum distributions of two
independent particles in the CMS of a small volume $dV$ (where they
are assumed to be spherically symmetrical) integrated with the factor 
$\delta[(p_1
+\,p_2)^2-\,M^2]$, whereas the denominator accounts the normalization.}.

3. To compare the above approach straightforwardly with the experimental data on
dilepton production, one should to link the numbers of pions and valons within
fireball and the number of the observed charged particles. This relation is
suggested to be as follows:\begin{equation}
N_{ch}\simeq\,\frac{2}{3}N_{\pi}+\,2N_{\bar Q}+\,0,4\,N_B\end{equation} where
$$N_B=\,\frac{1}{3}(\lambda^2\,-\,1)N^{u+d}_{\bar
Q}\simeq\,\frac{1}{3}(\lambda^2\,-\,1)N_{\bar Q}$$ is the fireball total baryon
number and, thus, 0,4$\,N_B$ is approximately equal to the proton
outcome from the fireball (in the accordance with the approximate ratio of
protons and neutrons in the heavy ion collisions). Eq.(8) implies also that,
in course of {\it thermally equilibrium} hadronization, the
number of decoupled pions emerged from the fireball is nearly equal to the
number of pions which were coupled within it and that each $Q\bar Q$ pair
produces about 2 charged pions (actually, about three of them, the third being
neutral).

Now, combining  eq.s (5 - 8), we come to the basic formula for the
excess in dilepton yield which is to be compared to the observations:  $$
\frac{1}{N_{ch}}\frac{dN_{ee}}{dM}\,\simeq\,
\frac{0,1\,\lambda^2\,(1\,+\,0,6\lambda\,+\lambda^2)^{-1}}
{4,7\,+\,\lambda\,+\,0,33 \lambda^2}\,\,\frac{\tau}{\bar t}\,\,
[\,\frac{dW_{\pi\pi}}{dM}\frac{\sigma_{\pi\pi\to
ee}}{\sigma_{\pi\pi}^{tot}}\,+ \,4\,\frac{dW_{Q\bar
Q}}{dM}\,\frac{\sigma_{QQ\to ee}}{\sigma_{Q\bar Q}^{tot}}$$
\begin{equation}+\,\beta\,\times\,(the\,\, previous\,\, item,\,\, where\,\,
Q_s\,\, stands\,\, instead\, \, of\,\,Q\,\equiv\,Q_{u,d})\,],
\end{equation} where $\sigma_{Q\bar Q}$ stands for the half-sum of the cross
sections of $Q_u \bar {Q}_u$ and of $Q_d \bar {Q}_d$ annihilation into $e^+ e^-
$,
$\sigma_{\pi\pi}^{tot}$ and $\sigma_{Q\bar Q}^{tot}$ stand for the
corresponding total cross sections, and $\beta$ accounts the
relative rate of strange valon collisions, $\beta\,\simeq$ 0,1 (see above).
The relative effectiveness of the latter ones in dilepton production
is still $\simeq$ 1,7 times lower\footnote{This coefficient is resulted as the
interplay of two factors: the ratio of quark electromagnetic charges squared,
$(e^2_u\,+\,e^2_d)/2\,e^2_s\,=\,2,5$, and the ratio $\sigma^{tot}_{Q\bar
Q}/\sigma^{tot}_{Q_s\bar {Q}_s}$ estimated from comparison of $\pi p$ and $Kp$
cross sections to be nearly equal to 1,5.}, thus being
about 6 \% of the light quark one. That is why below the strange valon
contribution is neglected. Of course, eq.(9) could be reformulated to look more
traditionally (for the latter see \cite{Gale}),
making use the above relations and well known definition $\sigma^{tot}_i \bar
{t}_i sqrt{2}\simeq\,n_i^{-1}$, where $n_i$ is the number density of $i$th
particle species and $\sigma^{tot}_i$ and $\bar t_i$ are the relevant cross
section and mean free time, respectively. The chosen way of description is
preferred here deliberately to avoid unnecessary complications. Moreover,
its physical meaning seems more transparent, whereas the drawbacks of
two ways of description are essentially equivalent (poorly defined $\pi \pi$ and
$Q\bar Q$ total cross sections in the one given here and equally poorly defined
corresponding in-medium $\rho$-meson electromagnetic formfactor which would
enter inevitably the usually exploited formulae).

Two general results can be deduced from eq.(9) right away, before going into
more detailed calculations. First, the mean number of successive interactions of
a particle over the fireball evolution time, $\tau/\bar t$, is the only factor 
on the right-hand side of eq.(9) which depends on $N_{ch}$:
$\tau\,\sim\,V\,\sim\,N_{ch}$ and $\tau\,\sim\,V^{1/3}\,\sim\,N^{1/3}_{ch}$ in
the limiting cases of one-dimensional (longitudinal) and three-dimensional
(spherical) expansion of the fireball, respectively ($V$ is the fireball volume
at the freeze-out temperature); thus, eq.(9) admits apparently the trend
$N_{ee}\,\sim\,N^2_{ch}$ observed in the SPS experiments \cite{js}.
Second,the predicted excess in dilepton yield is almost insensitive to the
choice of valon chemical potential within the range 0 $\leq\,\mu_Q\,\leq$ 80 
MeV,
since the corresponding variation of the $\lambda$-dependent factor on the
right hand side of eq.(9) does not exceed 20 \% \footnote{However, at lower
energies (AGS, SIS), when this potential is expected to be considerably higher,
its influence should result in quite observable (about 2 times) diminishing the
dilepton excess in collisions of the same nuclei as compared to the SPS one.
At the same time, it should be noted that similar effect comes from decrease
of the ratio $\tau/\bar t$ at too low energies, when the fireball initial
temperature $T_i\,<\,T_{ch}$.}.

The most vulnerable point of the formula (9) is the ratio of the above
cross sections. It can be estimated within the
framework of the following reasoning. The resonance irregularities (mostly,
the $\rho$-meson one) in the cross section of dilepton production and in
the total one are expected to be of the similar (the same within VDM) shape
\cite{Joss} and cancel each other substantially.  That is why the background
contributions (namely they are meant under the letter $\sigma$ below) are to be
compared only.  Thus, $$\sigma_{\pi\pi\to ee}\simeq\,\frac{4\pi \alpha ^2}{3
M^2} \quad \mbox{and} \quad \sigma_{Q\bar Q\to ee}\simeq\,\frac{10\pi \alpha
^2}{27 M^2}$$.  The total cross sections of $Q\bar Q$ and $\pi \pi$
interactions can be estimated, obviously, only in terms of plausibility and
similarity to the known hadronic ones.  We assume that, in spirit of the
suggested approach, each of $u$ and $d$ valons (antivalons) interacts
(strongly), as if it were "1/3 of the proton" (antiproton).  Since the cross
section of $p\bar p$ interaction \cite{pd} exhibits almost no resonance
structure and can be fitted pretty well (at not too high CMS energies $E_c$) by
the simple phenomenological formula\footnote{This formula corrected by
multiplication of its right-hand side by the factor $(E_c/2m_p)^{-0,4}$ (which
is of no importance within the energy range of our interest here) smoothly
interpolates between the threshold behavior of the total $p\bar p$ cross
section and its high energy parametrization \cite{pd} motivated by the Regge
approach.}\begin{equation} (\,\sigma_{p\bar p}(E_c)\,-
40\,\mbox{mb})\,\simeq\,\frac{24\,\mbox{mb}}{\sqrt{\frac{E_c}{2\,m_p}-\,1}},
\end{equation}
one can expect that the corresponding cross section of the light
valon-antivalon interaction respects approximately the formula \begin{equation}
(\,\sigma_{Q\bar
Q}(M)-\,9\,\mbox{mb})\,\simeq\,\frac{5,4\,\mbox{mb}}{\sqrt{\frac{M}{2\,m_Q}-
\,1}}
\end{equation}
where the $Q_{u,d}\bar Q_{u,d}$ cross section at the CMS energy $M\,=\,E_{c}/3$
is estimated to be about 0,22 (instead of 1/9) of the $p\bar p$ one at the
CMS energy $E_c$, taking into account $\simeq$ 50 \% shadowing in the $p\bar p$
interaction\footnote{This qualitative estimate is emerged from the observation
that $\pi p$ and $pp$ total cross sections at the relevant energies are more
likely related as 5:6 \cite{pd}, than as 2:3, what would be expected, if no
shadowing were at all. By the way, eq.(11) is coordinated pretty good with the
(also qualitative) estimate that $Q\bar Q$ total cross section at CMS energy of
several GeV should be of the order of confinement radius squared, i.e.,
$\simeq$ 1\,fm$^2$ $\simeq$ 10 mb.}.

This way of reasoning is not applicable
immediately for linking the shapes of the $p\bar p$ and $\pi^+ \pi^-$ total
cross sections, first of all, because the former reaction is
exothermic\footnote{That is why the simplest interpolation of the form
$\sim\,(E_c\,-\,2\,m_p)^{-\frac{1}{2}}$ fits its cross section fairly well.}
(just like the $Q\bar Q$ one), whereas the latter one is not, and therefore,
the above cross sections differ essentially in their threshold behavior.
Nevertheless, the available experimental data show up unambiguously a general
trend of all the known hadronic cross sections (apart from their resonance
structure, i.e., averaged with respect to it - namely, of their background
components which are just relevant) to increase gradually toward the threshold
(except of, maybe, a very narrow domain in the close vicinity of the
threshold). Compiling the data, one can conclude that these cross sections show
up a (2\,$\div$\,3) time decrease as the CMS energy increases by about 1 GeV
above the threshold, and that they approach the almost constant values as the
energy exceeds substantially the sum of the interacting particle masses.  We
seem reasonable to assume that the same is qualitatively true for the
background component of the $\pi^+ \pi^-$ cross section as well.  That is why
the following formula can be proposed:  \begin{equation}(\,\sigma_{\pi
\pi}(M)\,-\,\sigma_0\,)\,\simeq\,\frac{3\,\sigma_0\Delta}{
\sqrt{\frac{M}{2\,m_{\pi}}\,-\,1}\,+\,\Delta}\end{equation}
where $\sigma_0\,\simeq$ (10 $\div$ 15) mb is the high energy value of
$\sigma_{\pi \pi}$ and $0,2\,\leq\,\Delta\,\leq 1$, a some low value of $\Delta$
from this interval seeming rather more suitable because of the relatively small
pion mass.  \\

4. After insertion of eq.s (11 - 13) into the basic eq.(9), we are almost ready
for comparing the approach with the data. What remains to be done is to
adapt the formula (7) to the specific conditions of the measurements, i.e. to
take into account that the leptons, $e^+$ and $e^-$, with transverse momenta
$p_T>\,200$ MeV have been selected only in all the data and that the further
selection of the data into two groups incorporating
the events with dilepton total transverse momenta 200 MeV $<\,q_T<$ 500 MeV
and $q_T >$ 500 MeV, respectively, has been made.  These restrictions are
allowed for approximately: the limits of
integration over $\xi$ in eq.(7) are chosen to account the above conditions in
the average (i.e., these limits correspond to a "typical dilepton" built up of 
two leptons, whose momenta are averaged
over their relative directions and over their absolute values). As a result,
$$max\,[\,M,\,p_T\sqrt{6}\,\simeq\,0,5\, \mbox{GeV}\,], \quad p_T =\,0,2
\,\mbox{GeV},$$ stands instead of $M$ for the lower limit and
$$\sqrt{M^2+\,\frac{3}{2}q_T^2},\quad q_T=\,0,5 \,\mbox{GeV},$$ stands for the
upper (lower) limit for the events with $q_T<\,0,5$ GeV \, ($q_T>\,0,5$ GeV).\\

Below, the results of the suggested approach are compared to the experimental
data and to the theoretical predictions obtained within the
frameworks of some conventional theoretical approaches.\\

DISCUSSION AND CONCLUDING REMARKS.\\

At first sight, a doubt could be expressed that the valons play an essential
role in producing the dileptons with masses $M\,\leq\,2 m_Q\,\simeq$ 660 MeV
which are just of the primary interest here, since they do not produce these
dileptons directly.  However, it is not correct: they do play this role because
they directly affect the number of pions within the fireball and the number of
charged particles in the final state, and, therefore, their influence on the
ratio of dileptons to charged particles is quite unambiguous.

Unfortunately, we can not extract from our results the direct information about
the duration of fireball expansion $\tau$ and about the mean free time $\bar t$
separately because dilepton yield is proportional to their ratio only. The
curves presented in Fig's. 1 and 2a,b are obtained provided that $\tau/\bar
t$ = 20 or 30 for free pions ($m_{\pi}$ = 140 MeV) or "in-medium" pions
($m_{\pi}$ = 100 MeV), respectively. At the same time, the quadratic growth of
the total dilepton yield as compared to the charged particle one,
$N_{ee}\,\sim\,N^2_{ch}$, that was drawn \cite{js} from the CERES data witnesses
in favor of the predominantly longitudinal fireball expansion at the SPS
energies. The well known estimates predict in this case a rather long
expansion time \cite{Cher,Sto}. If one adopts a reasonable estimate for the mean
free time, $\bar t\,\simeq\,(0,7\,\div\,1)$ fm, then the duration of
pion-valonic phase (which is nearly equal to the entire duration of fireball
expansion) is $\tau\,\simeq\,(15\,\div\,20)$ fm or (20 $\div$ 30) fm for free 
and in-medium pions, respectively. These values are compatible with what was
predicted, the latter seeming somewhat more preferable.

At the
same time, due to a considerable enhancement of pressure within the fireball
along with increase of its energy density predicted in the framework of the
hydrodynamical model, the role of transverse flow increases too, and thus, the
three-dimensional pattern of fireball expansion is expected to become 
substantially
more pronounced in RHIC and LHC collisions. As a result, a certain
modification in the functional form of the above correlation is expected to be
observed, the 2th power there being gradually decreased toward the 4/3th one
(which refers to the spherical expansion). Thus, at RHIC, one can expect a
noticeably slower rise of the excess in dilepton yield over the CERES
cocktail sample, than it would be predicted by careless extrapolation of
the longitudinal expansion models (say, than $\simeq$ 4,5 times as compared to
SPS, if one assumes that $N_{ch}\,\sim\,E_c^{1/2}$).

Within the suggested approach, the three-particle collisions were neglected.
This approximation can be justified, only if the mean interaction time
($\simeq$ the particle size) is much smaller, than the mean free time ($\simeq$
the mean free path). One has to agree that the above estimate of reasonable
value of the mean free time, $\bar t\,\simeq\,(0,7\,\div\,1,0)$ fm,
does not answer this requirement completely even for the valons
of the size about 0,3 fm. A typical diagram that refers to three-particle
collision is shown in Fig. 3. Intervention of
a third particle (labelled as 3) results, apparently, in some dispersion of
dilepton mass around its only value which would be prescribed by the energy
conservation low, if only two particles were collide. In turn, it results in
smoothing the irregularities (kinks, dips or bumps), if they are inherent in
the dilepton mass spectrum predicted by two-particle collision kinematics.  In
particular, the dip in the spectrum before the two-pion threshold obtained in
the two-particle collision approximation, see Fig. 2a, could be flatten, to
some extent, by this smoothing. However, the relevant corrections are hardly
sufficient to level this dip completely - we mean a quite probable minimum also
seen there in the data specially selected to emphasize the contribution of
pion-pion collisions.

At the same time, Fig. 1 shows that occurrence of a
noticeable dip before $2m_{\pi}$ threshold is apparently predicted in
the entire bulk of data, if the in-medium pion mass equals to the free one,
whereas almost no visible dip is predicted, if an effective
lowering of this mass is qualitatively allowed for by taking it
equal to 100 MeV.  That is why the essential refinement of the data within this
mass region is asked anxiously because it can provide the valuable information
on properties of dense and hot matter that might occur even more important,
than the dilepton yield itself. In particular, it can be correlated to the
properties of the chiral transition.  As for the properly selected events,
$q_T\,\leq$ 500 MeV (see Fig. 2), we would like to point out again that this
dip is quite apparently predicted by either of the above versions of the
theoretical approach suggested, in contrast to the conventional ones.  Again
the allowance for a some decrease of the in-medium pion mass seems fruitful
(although three particle interaction could be responsible too (see above) for a
slight shift to the left of the minimum suggested by the data from the position
predicted by the free pion mass version of the approach presented).

The predicted yield of dileptons with $M_{\rho}\leq\,M\,\leq\,M_{\phi}$ looks
slightly overestimated, see Fig.s 1, 2.  However, this excess
(if it can be taken seriously into
consideration in face of too poor accuracy of the data)
which is undoubtedly due to $Q\bar Q$ annihilation may occur rather
illusive: in particular, at the $\phi$-meson peak, the experimental points are
situated even slightly lower, than they are expected to occur according to the
estimate of the prompt resonance (CERES cocktail) contribution itself, what
seems unreasonable.  This disparity can be taken as a hint that something here
may suffer from a systematical error.  If it is the case, then the agreement
between our predictions and the data is improved irrespectively of what -
data or CERES cocktail - is to be corrected, since what we have calculated is
just the expected excess in the observed dilepton yield over the CERES cocktail
sample.  At still higher dilepton masses, both the data and our results (unlike
the other ones) show up
the quite compatible excess over this sample, see Fig. 2b, which is undoubtedly
due to $Q\bar Q$ annihilation.

An advantage of the suggested approach is that its
physical sense and internal structure are very transparent and opened for
discussion, tracing and corrections, since no complicated generators are
involved: almost all the calculations are quite simple for being performed
approximately by hands.

Summing up what was said above, we conclude that the
low-mass spectrum of dileptons produced in course of the heavy ion collisions
can be understood in quite natural way in terms of the pion-valonic contents of
the expanding hot and dense matter (fireball) below the chiral transition
temperature. The above consideration showed, however, that dilepton production
is affected by a number of factors and that some of them could be estimated
rather semi-quantitatively.  Thus, we are supplied again with an insight on
fruitfulness of using the notion of valon, although the suggested evidences in
favor of its right to be acknowledged as a real physical object are still far
from being decisive.

The authors are indebted to N.G. Polukhina for the valuable help in
composing the figures.

FIGURE CAPTIONS\\

Fig. 1.  Our results (bold solid and bold dashed lines refer to $m_{\pi}$ = 140
MeV, $\tau/\bar t$ = 20 and $m_{\pi}$ = 100 MeV, $\tau/\bar t$ = 30,
respectively; $\sigma_0$ = 10 mb, $\Delta$ = 0,2 GeV) are confronted to the
entire bulk of CERES dilepton data \cite{exp} and predictions of thermal
dilepton calculations quoted from ref.\cite{Rapp,Rap}

Fig. 2.  The same as in Fig. 1 for the two $q_T$ selected groups of the data.

Fig. 3.  A typical diagram of three-particle collision. It shows that
intervention of the 'third" particle can affect the mass of produced dilepton
pair to shift it up or down.  In particular, this mass can occur below the
physical threshold of the corresponding two-particle (1 + 2) reaction. A very
likely minimum before the two-pion threshold seen in Fig. 2a is indicative for
the conclusion that contribution of three-particle collisions to dilepton
yield is rather small.

\end{document}